\documentclass[%
 reprint,
superscriptaddress,
amsmath,amssymb,
aps,
floatfix,
]{revtex4-2}

\usepackage{graphicx}% Include Figure~files
\usepackage{dcolumn}% Align table columns on decimal point
\usepackage{bm}% bold math
\usepackage{amsmath,amssymb}

\begin{document}

\preprint{APS/123-QED}

\title{Robust atomic orbital in a cluster magnet, LiMoO$_2$}

\author{N. Katayama}\email{katayama.naoyuki@b.mbox.nagoya-u.ac.jp}
\affiliation{Department of Applied Physics, Nagoya University, Furo-cho, Chikusa-ku, Nagoya, Aichi 464-8603, Japan}
\author{H. Takeda}
\affiliation{Institute for Solid State Physics, University of Tokyo, Kashiwa, Chiba 277-8581, Japan}
\author{T. Yamaguchi}
\affiliation{Department of Physics, Chiba University, 1-33 Yayoi-cho, Inage-ku, Chiba 263-8522, Japan}
\author{Y. Yamada}
\affiliation{Department of Applied Physics, Nagoya University, Furo-cho, Chikusa-ku, Nagoya, Aichi 464-8603, Japan}
\author{K. Iida}
\affiliation{Neutron Science and Technology Center, Comprehensive Research Organization for Science and Society (CROSS), Tokai, Ibaraki 319-1106, Japan}
\author{M. Takigawa}
\affiliation{Institute for Solid State Physics, University of Tokyo, Kashiwa, Chiba 277-8581, Japan}
\author{Y. Ohta}
\affiliation{Department of Physics, Chiba University, 1-33 Yayoi-cho, Inage-ku, Chiba 263-8522, Japan}
\author{H. Sawa}					
\affiliation{Department of Applied Physics, Nagoya University, Furo-cho, Chikusa-ku, Nagoya, Aichi 464-8603, Japan}
\date{\today}

\begin{abstract}
In this study, we present a rutile-related material LiMoO$_2$ becomes a cluster magnet and exhibits a spin singlet formation on a preformed molybdenum dimer upon cooling. Unlike ordinary cluster magnets, the atomic $d_{yz}$ orbital robustly survives despite the formation of molecular orbitals, thereby affecting the magnetic properties of the selected material. Such hybrid cluster magnets with the characters of molecular and atomic orbitals realize multiple independent spins on an isolated cluster, leading to an ideal platform to study the isolated spin dimers physics.
\end{abstract}

\maketitle

%\tableofcontents
Transition metal compounds with multiple electron degrees of freedom frequently form molecular orbitals in a spin singlet state \cite{Li2RuO3,CuIr2S4,LiRh2O4,MgTi2O4,Li033VS2,LiVS2,AlV2O4,LiVO2,CsW2O6-X,AlV2O4-2}. Examples include trimers of LiV$X_2$ ($X$ = O, S) \cite{LiVO2, LiVS2}, and dimers of VO$_2$, NbO$_2$ and MoO$_2$ \cite{VO2-1, VO2-2, VO2-3, VO2-4, NbO2-1, NbO2-2, Hiroi, MoO2_neutron}. Molecular orbital formation has been attracting a considerable attention because it occasionally accompanies a drastic metal-insulator transition \cite{Li2RuO3,CuIr2S4,LiRh2O4,MgTi2O4,LiVS2} and/or giant entropy changes \cite{Li033VS2,LiVS2,LiVO2}. This enables several possible applications in thermochromics \cite{thermo-1, thermo-2}, non-volatile memory \cite{memory}, and phase change materials \cite{Li033VS2, PCM-1, PCM-2, PCM-3, PCM-4}. Moreover, by arranging additional spins in the molecular orbital, exotic magnetic materials, called cluster magnets, can be obtained \cite{LiZn2Mo3O8-1, LiZn2Mo3O8-2, K2Mo8O16-1, K2Mo8O16-2, V4O7, BaV10O15-1, BaV10O15-2, Li2AMo3O8, TaS2-1, TaS2-2}.

LiZn$_2$Mo$_3$O$_8$ is an example of cluster magnets, where inherently formed trimer clusters of molybdenum ions collectively produce a spin $S$ = 1/2 moment \cite{LiZn2Mo3O8-1, LiZn2Mo3O8-2}. Due to the frustration effect between trimer clusters arranged in a triangular lattice, an exotic resonating valence bond state emerges at low temperature \cite{LiZn2Mo3O8-1, LiZn2Mo3O8-2}. The hollandite-type K$_2$Mo$_8$O$_{16}$ is another example, comprising molybdenum tetramer clusters with a spin $S$ = 1/2. Despite the strong antiferromagnetic interaction with the Weiss temperature of -34~K, the long-range ordering is suppressed to 2~K, expecting us a realization of spin-liquid like state \cite{K2Mo8O16-1, K2Mo8O16-2}. 

Rutile and its related compounds are ideal playground to explore novel cluster magnets because dimers frequently appear there and lithium intercalation is available to tune the electronic states. Lithium intercalation supplies electrons to the host structure, resulting in an additional spin to the preformed dimers, possibly leading to the development of novel cluster magnets. Here, we focus our attention on a $d^2$ compound, MoO$_2$, and its lithium intercalated derivatives. In the parent MoO$_2$, preformed dimers are formed, indicating that molecular orbitals are realized \cite{Hiroi, MoO2_neutron}. The preformed dimers are formed by a $\sigma$-bonding, while $\pi$-orbitals are strongly coupled with oxygen $p$-orbitals, resulting in a high metallic conductivity \cite{Hiroi, Moosburger}. Note that the schematics of atomic orbitals composing $\sigma$- and $\pi$-orbitals are shown in Figures~\ref{fig:Fig4}(a) and \ref{fig:Fig4}(b). Although details of the structural and physical properties of lithium intercalation compounds are yet to be clarified, the previous magnetic susceptibility measurements have shown that an anomalous temperature dependence appears in LiMoO$_2$ \cite{LiMoO2}, as shown in Figure~\ref{fig:Fig3}(a), expecting us an unusual magnetic property of a cluster magnet. 

Here, we first show that Li$_{0.5}$MoO$_2$ and LiMoO$_2$ display different patterns of preformed molybdenum dimers. While Li$_{0.5}$MoO$_2$ is a paramagnetic metal, LiMoO$_2$ becomes a Mott insulator with two localized electrons on a molybdenum preformed dimer. Based on our structural and theoretical studies, we conclude that the molybdenum preformed dimers comprise the multiple bonding of $\sigma$- and $\pi$-bonds, while the $\delta$-bond is not formed, thereby resulting in the localized $d$ electron in original $d_{yz}$ orbitals, as shown in Figure~\ref{fig:Fig4}(e). On cooling, the two spins on a molybdenum dimer form a spin singlet ground state. Therefore, LiMoO$_2$ is an unusual type of magnetic cluster where the original $d_{yz}$ orbitals robustly survive on preformed molybdenum dimers.

Powder samples of LiMoO$_2$ were prepared by a soft-chemical method. Commercial MoO$_2$ powder samples were immersed on 0.2~M $n$-BuLi hexane solution for two days to attain the maximum Li content. By decreasing the $n$-BuLi concentration, we successfully obtained Li$_{0.5}$MoO$_2$. The Li content of Li$_{0.5}$MoO$_2$ and LiMoO$_2$ was estimated to be approximately 0.5 and 1.0 respectively, by referring to the lattice parameters presented in a previous study \cite{LixMoO2}. For LiMoO$_2$, Li content was also estimated to be 0.98(2) from the Rietveld analysis of synchrotron x-ray diffraction experiments. Details of other experimental methods are summarized in Supplemental Information \cite{SI}.

\begin{figure}
\includegraphics[width=80mm]{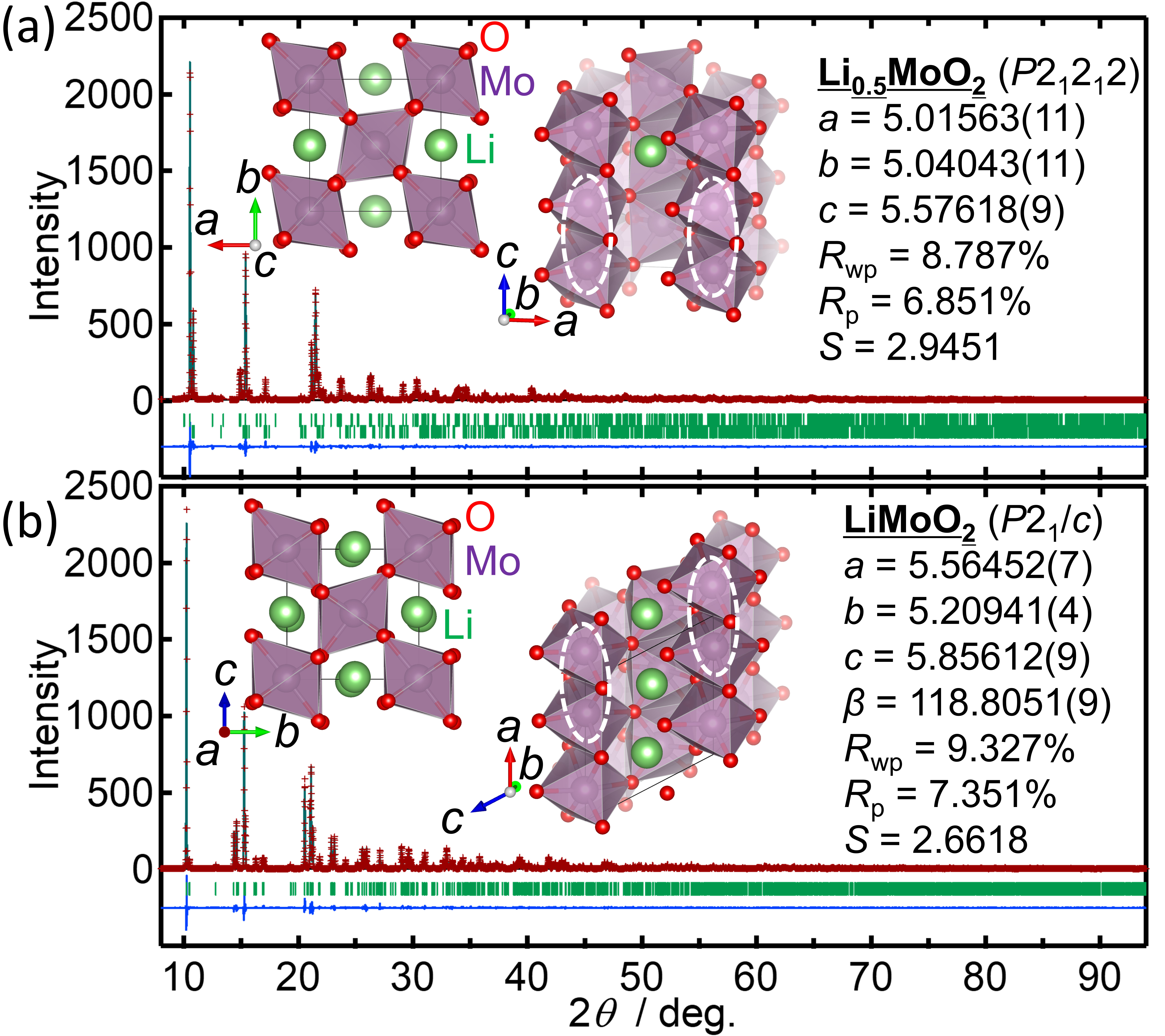}% Here is how to import EPS art
\caption{\label{fig:Fig1} Rietveld analysis results of (a) Li$_{0.5}$MoO$_2$ and (b) LiMoO$_2$, obtained at room temperature. In Li$_{0.5}$MoO$_2$ data, $\sim$27\% of Li$_{0.2}$MoO$_2$ with a space group $P$$2_1$/$c$ was identified as an impurity. Insets indicate the crystal structures. Dotted ovals indicate the preformed molybdenum dimers. Note that similar dimer patterns are realized in LiMoO$_2$ and MoO$_2$.}
\end{figure}

Figures~\ref{fig:Fig1}(a-b) show the Rietveld analysis of Li$_{0.5}$MoO$_2$ and LiMoO$_2$. Rietveld analysis of MoO$_2$ is shown in Supplemental Information \cite{SI}. While MoO$_2$ and LiMoO$_2$ were successfully refined by assuming the same monoclinic space group $P2_1$/$c$, an orthorhombic space group $P$2$_1$2$_1$2 was used to refine Li$_{0.5}$MoO$_2$. While the refined crystal structures of MoO$_2$ and LiMoO$_2$ were consistent with those reported before \cite{structure_LiMoO2}, the crystal structure of Li$_{0.5}$MoO$_2$ was first identified in this study. Although the emergence of preformed molybdenum dimers was identified in all the materials measured, the dimer arrangements of Li$_{0.5}$MoO$_2$ were different from those of MoO$_2$ and LiMoO$_2$. This indicates that the dimer arrangement was changed twice during the lithium intercalation. In Li$_{0.5}$MoO$_2$, the molybdenum dimers are located away from the intercalated lithium ions probably due to the Coulomb repulsion between periodically inserted lithium ions and molybdenum dimers. After multiple structural phase transitions, similar dimer arrangements were restored in LiMoO$_2$ with MoO$_2$. The refined structural parameters are presented in the Supplemental Information \cite{SI}.

\begin{figure}
\includegraphics[width=80mm]{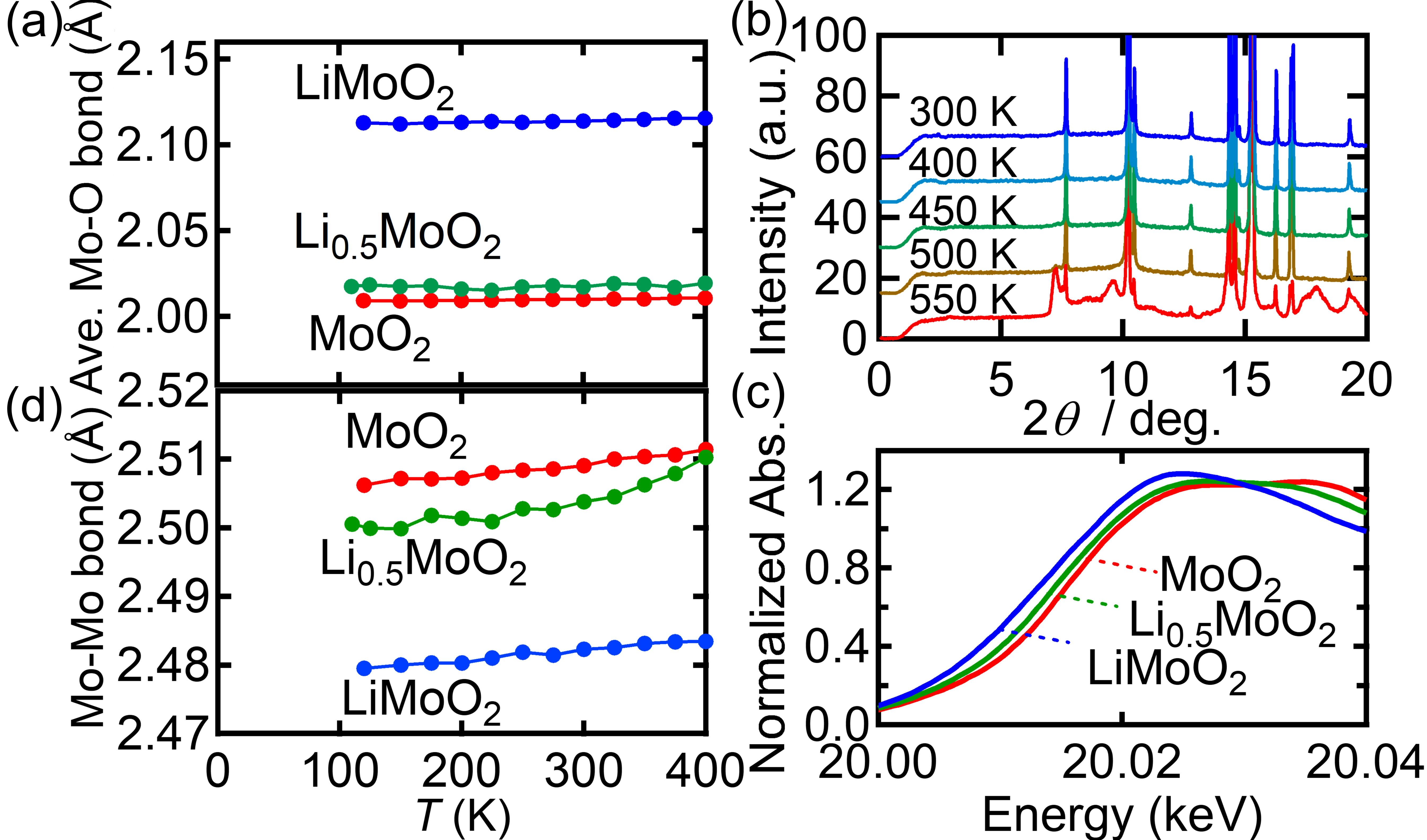}% Here is how to import EPS art
\caption{\label{fig:Fig2} (a) Average of the six Mo-O distances constituting of the MoO$_6 $ octahedron. (b) Temperature dependence of x-ray diffraction patterns of LiMoO$_2$. (c) XANES spectra for MoO$_2$, Li$_{0.5}$MoO$_2$ and LiMoO$_2$. (d) Mo-Mo distances consisting of preformed dimers.}
\end{figure}

While the preformed molybdenum dimers commonly appear in these compounds, considerable differences exist in the local structures. Figure~\ref{fig:Fig2}(a) shows the averaged Mo-O distances obtained from the six Mo-O bonds constituting the MoO$_6$ octahedron. Note that the averaged Mo-O distance is kept almost constant up to the decomposition temperature of $\sim$550~K for LiMoO$_2$, as shown in Figure~\ref{fig:Fig2}(b). Compared with MoO$_2$ and Li$_{0.5}$MoO$_2$, the averaged Mo-O distance is $\sim$0.1~\AA~longer for LiMoO$_2$.  Considering lithium intercalation inevitably changes the formal valence of molybdenum from the 4+ to 3+ side, one may consider that the increase in the ionic radius of molybdenum ions is responsible for the elongation. Note that the change in the formal valence of molybdenum caused by the lithium intercalation was confirmed from the molybdenum K-edge EXAFS experiment, as shown in Figure~\ref{fig:Fig2}(c). The edge energy is shifted to the lower energy side in the order of MoO$_2$-Li$_{0.5}$MoO$_2$-LiMoO$_2$, indicating that the formal valence of molybdenum changes from the 4+ to 3+ side via lithium intercalation. However, according to Shannon Ionic Radii \cite{Shannon}, the ionic radii of octahedrally coordinated Mo$^{3+}$ and Mo$^{4+}$ are 0.69~\AA~and 0.65~\AA, respectively. The difference, 0.04~\AA, is very small to explain why the change of the averaged Mo-O distance reaches $\sim$0.1~\AA. Therefore, we expect that the Mo4$d$-O2$p$ orbital hybridization becomes weaker in LiMoO$_2$ compared with MoO$_2$ and Li$_{0.5}$MoO$_2$. Conversely, the Mo-Mo distance of preformed dimers becomes 0.03 \AA~shorter in LiMoO$_2$ compared with MoO$_2$, as shown in Figure~\ref{fig:Fig2}(d), indicating that the molybdenum dimers are enhanced in LiMoO$_2$ compared with MoO$_2$.

Next, we revisit the physical properties of these samples. These results are basically consistent with the previous data \cite{LiMoO2, LiMoO2-2}. An almost temperature-independent Pauli paramagnetism was observed for MoO$_2$ and Li$_{0.5}$MoO$_2$ from the magnetic susceptibility experiment \cite{LiMoO2}. Correspondingly, the low temperature sintered samples of MoO$_2$ and Li$_{0.5}$MoO$_2$ show a metallic conductivity down to the lowest temperature measured \cite{LiMoO2-2}. However, a strong temperature dependence appears in LiMoO$_2$, as shown in Figure~\ref{fig:Fig3}(a). The observed behaviors are qualitatively consistent to those previously reported \cite{LiMoO2}, although the origin of the strong temperature dependence on LiMoO$_2$ is yet to be clarified. When the band structure near the Fermi surface has a characteristic shape, temperature dependent behavior occasionally appears even in a Pauli paramagnetic metal. However, our low temperature sintered sample of LiMoO$_2$ showed no electrical conductivity at room temperature \cite{LiMoO2}. This suggests that LiMoO$_2$ is essentially an insulator with a large charge gap and the temperature dependence of the magnetic susceptibility is not derived from the Pauli paramagnetism.

\begin{figure}
\includegraphics[width=80mm]{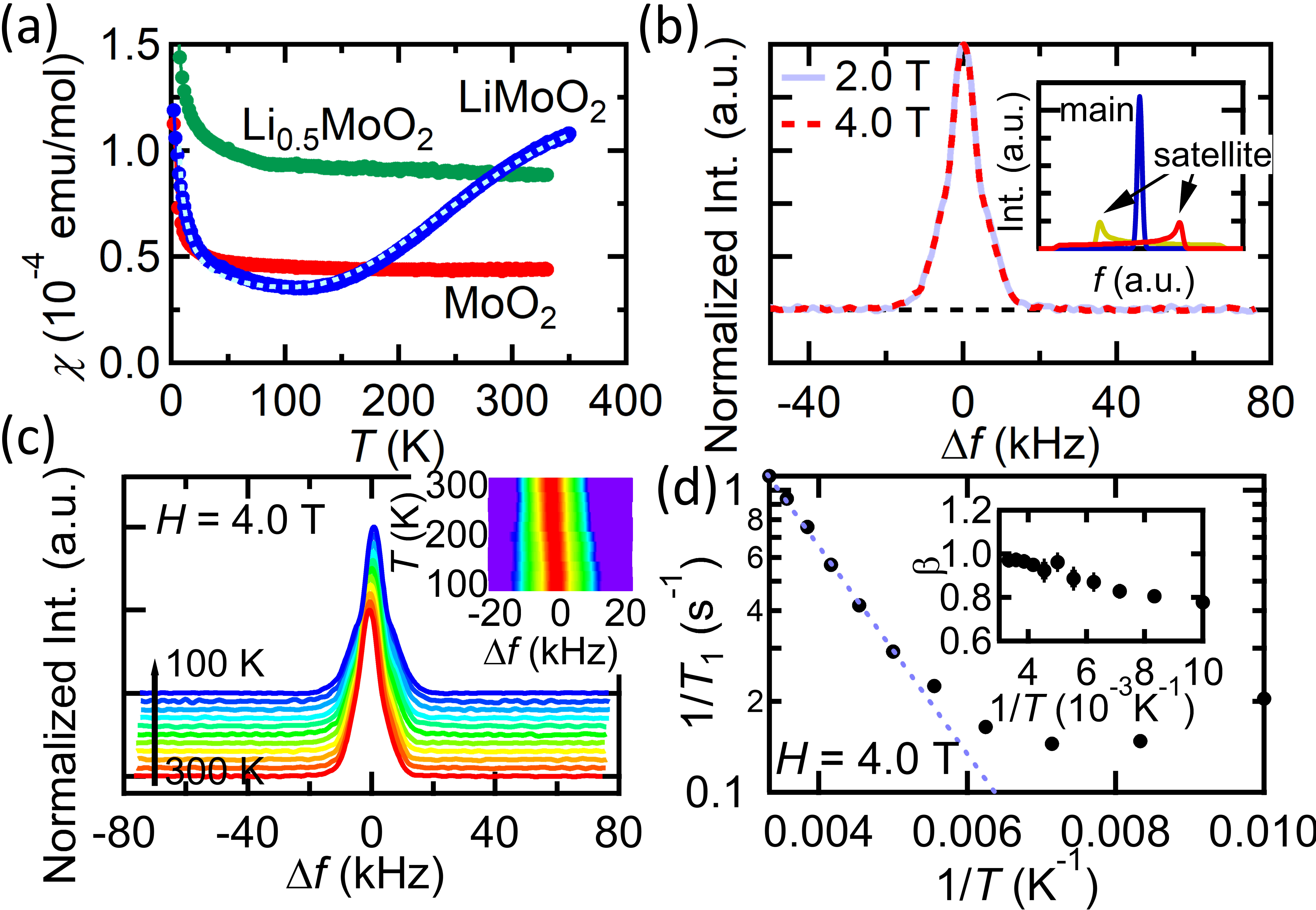}% Here is how to import EPS art
\caption{\label{fig:Fig3} (a) Magnetic susceptibility for MoO$_2$, Li$_{0.5}$MoO$_2$ and LiMoO$_2$. $H$ = 1~T was applied for collecting the data. For LiMoO$_2$, the data was fitted by the white dotted line using the equation for isolated dimers \cite{Bleaney}. (b) $^7$Li-NMR spectra obtained at 200~K for mangetic fields of 2.0 and 4.0~T for LiMoO$_2$. $\Delta$$f$ is defined as the frequency shift measured from the reference frequency, $\Delta$$f$ = $f$-$\gamma$$H$. Inset shows the simulated pattern of typical powder patterns for -3/2 $\leftrightarrow$ -1/2 (red), 3/2 $\leftrightarrow$ 1/2 (yellow) and -1/2 $\leftrightarrow$ 1/2 (blue) when $I$ = 3/2. (c) Temperature dependences on $^7$Li-NMR spectra for LiMoO$_2$. Inset shows the contour map of the intensity at various temperatures. (d) Spin-lattice relaxation rate (1/$T_1$) and the Arrhenius plot with the function of 1/$T_1$~=~$a$~exp(-$\Delta$/$T$) for LiMoO$_2$. The inset shows the temperature dependence of the stretch exponent, $\beta$.}
\end{figure}

Because LiMoO$_2$ is inherently an insulator, we can expect that the antiferromagnetic or spin singlet state should be realized. To identify the magnetic ground state, $^7$Li-NMR measurement of LiMoO$_2$ was performed. The NMR spectrum at 200 K displayed in Figure~\ref{fig:Fig3}(b) shows a relatively sharp central peak on top of a slightly broader line. The latter corresponds to the quadrupole satellite ($\pm$3/2 $\leftrightarrow$ $\pm$1/2) transitions, although the line shape is quite different from the simulated typical powder pattern shown in the inset. This is because the quadrupole shift is very small, comparable with the width of the central line which is likely to be determined by field independent nuclear dipole-dipole interactions. As shown in Figure~\ref{fig:Fig3}(c) and its inset, the linewidth of the main peak is unchanged when the temperature is reduced. These results show that the ground state is not antiferromagnetic. The line width of the satellite line slightly broadens with decreasing temperature, which may be due to reduced mobility of lithium ions. The spin lattice relaxation rate, 1/$T_1$, follows an activation law as shown in Figure~\ref{fig:Fig3}(d), indicating the emergence of a spin gap at low temperature. The size of the spin gap was estimated to be approximately 800~K from the Arrhenius plot. The 1/$T_1$ deviates from the activation law below $\sim$180~K, accompanied by moderate decrease of stretch exponent $\beta$ from unity. This is ascribed to the appearance of some inhomogeneous spin relaxation process such as the relaxation via impurity spins or lithium motion.

As presented above, our experimental findings clearly indicate that LiMoO$_2$ is essentially an insulator with a spin gap. Next, we discuss the origin of the spin gap from the structural aspects. Compared with MoO$_2$, the intra-dimer distance becomes shorter in LiMoO$_2$. This indicates that the preformed dimer is enhanced in LiMoO$_2$. A possible scenario is that hybridization of the $\pi$ molecular orbital and O2$p$ orbital become weak in LiMoO$_2$, while the $\pi$ bond strengthens the molybdenum preformed dimer. That is, the multiple bonds comprising $\sigma$ and $\pi$ bonds are realized in LiMoO$_2$, as shown in Figures~\ref{fig:Fig4}(a-b). This bonding state is distinctly different from that in MoO$_2$, where the $\pi$ molecular orbital is strongly hybridized with the O2$p$ orbital, resulting in strong electronic conductivity. Compared with MoO$_2$, the averaged Mo-O distance is apparently longer in LiMoO$_2$, suggesting weak hybridization of the $\pi$ molecular orbital and O2$p$ orbital. 

LiMoO$_2$ has $d^3$ electrons, and two of the three electrons contribute to the formation of $\sigma$ and $\pi$ bonds. Therefore, the remaining third electron should produce the spin gap. Two possible scenarios exist on the role of third electron. 

The first is that the third electron forms a $\delta$-bond, as shown in Figure~\ref{fig:Fig4}(c), resulting in LiMoO$_2$ becoming a band insulator with a narrow gap between the $\delta$- and $\delta^*$-orbitals, as shown in Figure~\ref{fig:Fig4}(d). When the gap between $\delta$- and $\delta^*$-orbitals is narrow, electrons in $\delta$-orbitals can be thermally excited into $\delta^*$-orbitals. The increasing numbers of thermally excited isolated spins should produce strong temperature dependent magnetic susceptibility on heating. However, this scenario does not seem to happen. From the $^7$Li-NMR experiment, a spin gap, relating to the singlet-triplet excitation, was estimated to be approximately 800~K. If LiMoO$_2$ is a narrow-gap band insulator with a gap of approximately 800~K, charge gap should be consistent with the spin gap. Therefore, many carriers should be thermally excited across the charge gap even at room temperature, leading to a high electrical conductivity. However, this critically contradicts our results. The charge gap should be considerably larger than the spin gap. 

The second is that the $\delta$-bonding is absent in LiMoO$_2$, while the third electron locates at the robustly surviving $d_{yz}$ orbital, as shown in Figure~\ref{fig:Fig4}(e). In this scenario, LiMoO$_2$ is inherently a Mott insulator with the third electron localized at each molybdenum site with a spin $S$ = 1/2. On cooling, spin singlet is formed between the neighboring third electrons localized on the preformed dimers. We can intuitively understand that this can happen because the overlap between adjacent $d_{yz}$ orbitals should be very small, as shown in Figure~\ref{fig:Fig4}(c), leading to the destabilization of $\delta$-bond formation. If the $d_{yz}$ orbitals robustly survives and form a dimer between localized spins, it is expected that the magnetic susceptibility data can be fitted using the equation of the isolated dimers. This fitting was successfully done using the equation below \cite{Bleaney},
\[
\chi  =  \frac{N g^2 \mu^2}{k_B T}  \frac{1}{3+e^{J/k_B T}}+\frac{C}{T}+B \nonumber
\]
where the second and third terms indicate the Curie impurity and temperature-independent Van Vleck terms with $\chi_{VV}$ = 2.98(3)$\times$10$^{-5}$ emu/mol. The estimated Curie constant of $C$ = 4.93(8)$\times$10$^{-4}$ emu/mol Mo$\cdot$K implies the existence of a tiny amount of paramagnetic impurities of less than 1 \% if we assume spin 1/2 moment. The effective magnetic moment is estimated to be $g\mu$ = 0.60(2)$\mu_B$, which is apparently smaller than the expected value. This possibly appears due to the weak hybridization between robust Mo 4$d_{yz}$ orbital and surrounding O 2$p$ orbitals, and/or spatially extended 4$d$ orbitals compared to 3$d$ orbitals. The fitting gives the great fit with $J$/$k_B$ $\sim$ 894 K, comparable to the estimated spin gap of $\sim$ 800 K from NMR experiment. The difference may be attributed to the narrow temperature window for fitting. This result supports that the present material can be treated as an isolated spin dimer model on a preformed dimer. The estimated spin gap of approximately 800~K seems to be unusually large for a spin $S$ = 1/2 system, possibly because the shortening of Mo-Mo distance caused by the preformed multiple bonds strengthens the exchange interaction between spins. In this scenario, the charge gap is expected to be considerably larger than the spin gap because LiMoO$_2$ is essentially a Mott insulator over the entire temperature range.

\begin{figure}
\includegraphics[width=80mm]{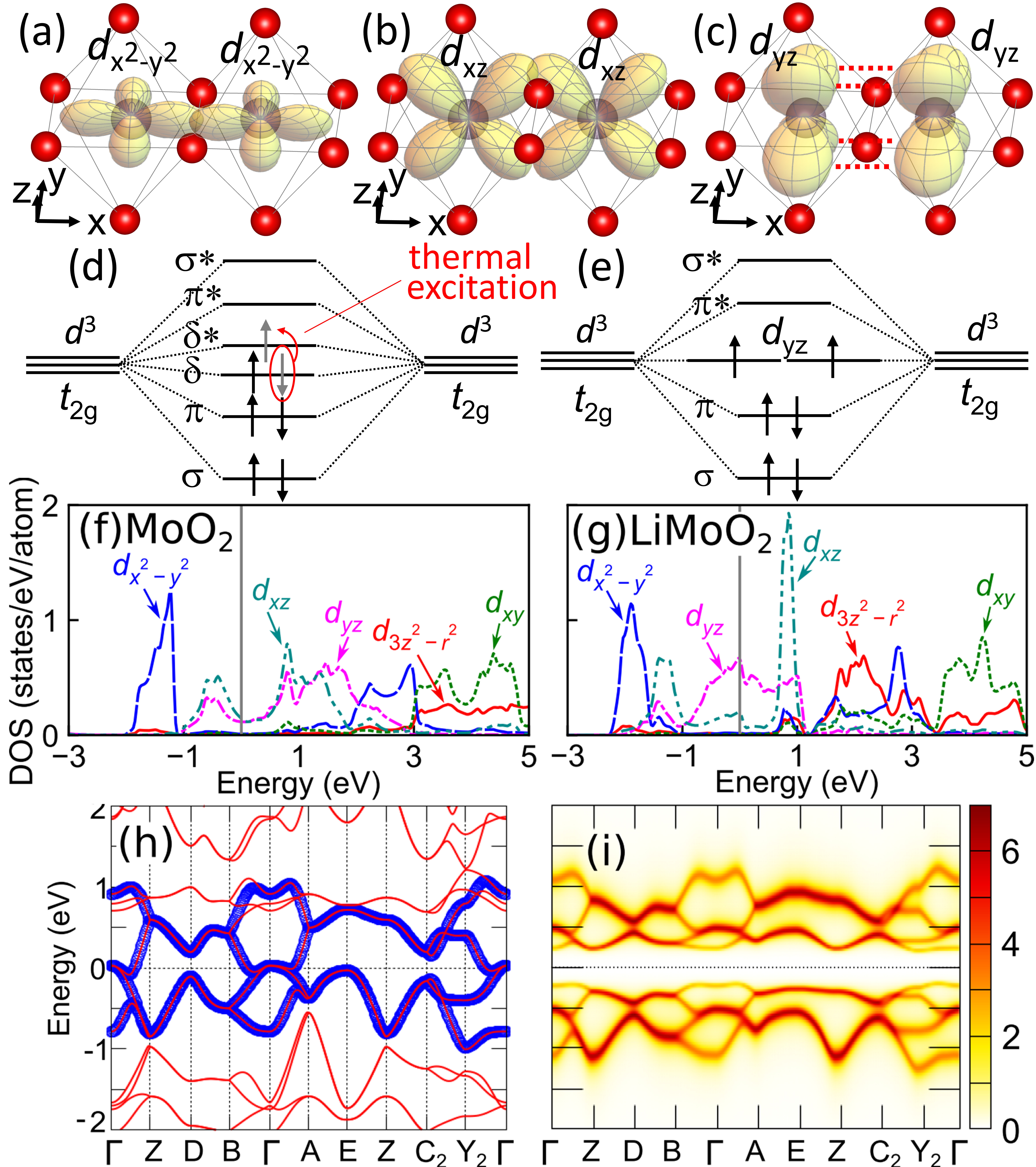}% Here is how to import EPS art
\caption{\label{fig:Fig4} The electronic structures of LiMoO$_2$ and MoO$_2$. (a)-(c) Schematics of the atomic orbitals composed of (a) $\sigma$-, (b) $\pi$-, and (c) $\delta$-bonds. (d-e) Schematics of the energy levels of the molecular orbitals of a molybdenum dimer where the gap opens between the $\delta$- and $\delta^{*}$-orbitals [(d)] but 
two nearly degenerate levels appear when the $\delta$-bonding is negligible [(e)]. (f-g) Calculated PDOSs for MoO$_2$ [(f)] and LiMoO$_2$ [(g)]. (h) Calculated band dispersions (red lines) compared with the tight-binding band dispersions obtained using four $d_{yz}$ MLWFs (blue lines). GGA ($U=0$ eV) is used for the calculations in (f-h). (i) Intensity plot of the single-particle spectral function calculated by the DMFT where we assume $U=1$~eV. }
\end{figure}

Second, we elaborate on the electronic states of the relevant materials from a theoretical perspective. Here, we applied the density functional theory (DFT) based on electronic structure calculations and the dynamical mean-field theory (DMFT) calculations \cite{Haule}. For the DFT calculations, we used the WIEN2k code \cite{WIEN2k} based on the full-potential linearized augmented-plane-wave method. We present calculated results obtained in the generalized gradient approximation (GGA) for electron correlations with the exchange-correlation potential of Ref.~\cite{perdew}. To improve the description of electron correlations in the Mo $4d$ orbitals, we used a rotationally invariant version of the GGA+$U$ method with double-counting correction in the fully localized limit \cite{anisimov, liechtenstein}.  In the self-consistent calculations, we typically use $\sim12\times12\times12$ ${\bm k}$-points in the full Brillouin zone.  Muffin-tin radii ($R_{\rm MT}$) of 1.54 (Li), 1.90 (Mo), and 1.74 (O) Bohr are used, and the plane-wave cutoff of 
$K_{\rm max}=7.00/R_{\rm MT}$ is assumed.  

The calculated results for the densities of states (DOSs) are shown in Figure~\ref{fig:Fig4}(f) for MoO$_2$ and in Figure~\ref{fig:Fig4}(g) for LiMoO$_2$. For MoO$_2$, we find that the partial DOS (PDOS) projected onto each $4d$ orbital of Mo ion corresponds perfectly with the previous work \cite{Moosburger}, where we note that, reflective of the strong $\sigma$-bonding nature, the bonding (antibonding) state of the $d_{x^2-y^2}$ orbitals is located $\sim$2 eV below (above) the Fermi level and the states from the $d_{xz}$ and $d_{yz}$ orbitals are located around the Fermi level, resulting in the metallic nature of the material. For LiMoO$_2$, we found that the bonding (antibonding) states of both the $d_{x^2-y^2}$ and $d_{xz}$ orbitals are located considerably below (above) the Fermi level, thereby allowing the assumption that the low-energy electronic properties are well described only by the $d_{yz}$ orbitals. Therefore, the standard DFT calculations predict that LiMoO$_2$ should be a metal, however, this contradicts the result of the experiment.

As shown in Figures~\ref{fig:Fig4}(d) and \ref{fig:Fig4}(e), the insulating behavior of LiMoO$_2$ can be explained by the band-insulator and Mott-insulator scenarios. In principles, the former may be justified by the DFT band-structure calculations. This is particularly the case when we introduce the on-site orbital potential $U$ in the GGA+$U$ scheme; however, we find that the band gap opens only if an unrealistically large value of $U\agt$~10~eV is introduced.  Therefore, we may conclude that the band-insulator scenario is highly unlikely.  

Because the Mott-insulator state cannot be described by the DFT calculations, we apply the DMFT calculations \cite{Haule}, using the DCore library \cite{DCore} based on the TRIQS \cite{TRIQS, DFTTool}. The Hubbard-I approximation is used for the impurity solver, which is known to provide successful results in well-localized systems \cite{Pourovskii, Amadon, Shick}. For the calculations, we first create a tight-binding band structure using four maximally localized Wannier functions (MLWFs) \cite{mostofi,kunes}, that corresponds well with the DFT band structure, as shown in Figure~\ref{fig:Fig4}(h), indicating that the low-energy electronic structure of this system is well described by the four $d_{yz}$ orbitals of Mo ions. The calculated result for the single-particle spectra at $U=1$ eV and $T\sim$~300~K in the DMFT is shown in Figure~\ref{fig:Fig4}(i) where we find that the band gap of a size $\sim$0.5~eV is clearly visible, indicating that the system is a Mott insulator. The gap size can be controlled by adjusting the value of $U$. We note that using an elaborative solver such as a continuous-time quantum Monte Carlo solver, or a self-consistent DFT+DMFT calculation, can facilitate more quantitative discussions on this material, however, this is beyond the scope of this paper.

Finally, let us verify the experimentally estimated spin gap from the theoretical point of view. Based on the tight-binding calculations results, we can estimate the hopping integral $t$ between nearest neighbor Mo ions to be $\sim$ 0.2 eV. Assuming that the experimentally obtained spin gap of 800-900 K, corresponding to 0.069-0.078 eV, is consistent to the exchange interaction $J$ based on the two-body model, we can estimate the on-site Coulomb interaction $U$ $\sim$ 2.06-2.32 eV through the function of $J$ $\sim$ 4$t^2$/$U$. This result is quite valid considering that LiMoO$_2$ is a 4$d$ electron system.

Thus, we can conclude that LiMoO$_2$ is a hybrid cluster magnet with the characters of molecular and atomic orbitals. Although robust atomic orbital is a unique feature of LiMoO$_2$ which cannot be found in ordinary cluster magnets, we can expect such electronic states can be realized on other cluster magnets as well when the outermost molecular orbital is unstable. This realizes an unusual situation where one cluster possesses multiple independent spins, provides an ideal platform to study isolated spin dimers physics.

\begin{acknowledgments}
The work leading to these results has received funding from the Grant in Aid for Scientific Research (Nos.~JP17K17793, JP17K05530,  JP25287083, JP18H04310 (J-Physics), JP17H02918, JP20K03829, JP20H02604 and JP19J10805) and Toyoaki Scholarship Foundation. T.Y. acknowledges support from the JSPS Research Fellowship for Young Scientists. This work was carried out under the Visiting Researcher’s Program of the Institute for Solid State Physics, the University of Tokyo, and the Collaborative Research Projects of Laboratory for Materials and Structures, Institute of Innovative Research, Tokyo Institute of Technology. The synchrotron powder x-ray diffraction and EXAFS experiments were conducted at the BL5S2 and BL11S2 of Aichi Synchrotron Radiation Center, Aichi Science and Technology Foundation, Aichi, Japan (Proposals No. 201702049, No. 201702101, No. 201703027, No. 201704027, No. 201704099, No. 201804016, No. 201904078 and No. 201704028), and at the BL02B2 and BL01B1 of SPring-8, Hyogo, Japan (Proposals No. 2019B1073 and 2017A1053). 
\end{acknowledgments}

\appendix

\nocite{*}

\bibliography{LiMoO2_references}% Produces the bibliography via BibTeX.

\end{document}